\documentclass[preprint,superscriptaddress,prl,amsmath,aps]{revtex4-1}

\usepackage{graphicx}
\usepackage{amsmath}
\usepackage{array}
\usepackage[mathlines]{lineno}
\makeatletter
\newcommand*{\rom}[1]{\expandafter\@slowromancap\romannumeral #1@}
\makeatother
\begin{document}

    \title{Frictional drag between superconducting LaAlO$_3$/SrTiO$_3$ nanowires}
	
	\author{Yuhe Tang}
    \affiliation{Department of Physics and Astronomy, University of Pittsburgh, Pittsburgh, Pennsylvania 15260, USA}
    \affiliation{Pittsburgh Quantum Institute, Pittsburgh, Pennsylvania 15260, USA}
    
    \author{Jung-Woo Lee}
    \affiliation{Department of Materials Science and Engineering, University of Wisconsin-Madison, Madison, Wisconsin 53706, USA}
    
    \author{Anthony Tylan-Tyler}
    \affiliation{Department of Physics and Astronomy, University of Pittsburgh, Pittsburgh, Pennsylvania 15260, USA}
    \affiliation{Pittsburgh Quantum Institute, Pittsburgh, Pennsylvania 15260, USA}
    
    \author{Hyungwoo Lee}
    \affiliation{Department of Materials Science and Engineering, University of Wisconsin-Madison, Madison, Wisconsin 53706, USA}
    
    \author{Michelle Tomczyk}
    \affiliation{Department of Physics and Astronomy, University of Pittsburgh, Pittsburgh, Pennsylvania 15260, USA}
    \affiliation{Pittsburgh Quantum Institute, Pittsburgh, Pennsylvania 15260, USA}
	
    \author{Mengchen Huang}
    \affiliation{Department of Physics and Astronomy, University of Pittsburgh, Pittsburgh, Pennsylvania 15260, USA}
    \affiliation{Pittsburgh Quantum Institute, Pittsburgh, Pennsylvania 15260, USA}
	
    \author{Chang-Beom Eom}
    \affiliation{Department of Materials Science and Engineering, University of Wisconsin-Madison, Madison, Wisconsin 53706, USA}
	
    \author{Patrick Irvin}
    \affiliation{Department of Physics and Astronomy, University of Pittsburgh, Pittsburgh, Pennsylvania 15260, USA}
    \affiliation{Pittsburgh Quantum Institute, Pittsburgh, Pennsylvania 15260, USA}
	
    \author{Jeremy Levy}
    \affiliation{Department of Physics and Astronomy, University of Pittsburgh, Pittsburgh, Pennsylvania 15260, USA}
    \affiliation{Pittsburgh Quantum Institute, Pittsburgh, Pennsylvania 15260, USA}
    \email[Corresponding author:]{jlevy@pitt.edu}
	
	\begin{abstract}

    We report frictional drag measurements between two superconducting LaAlO$_3$/SrTiO$_3$ nanowires. In these experiments, current passing through one nanowire induces a voltage across a nearby electrically isolated nanowire. 
The frictional drag signal contains both symmetric and antisymmetric components. The antisymmetric component arises from the rectification of quantum shot noise in the drive nanowire by the broken symmetry in the drag nanowire. The symmetric component in the drag resistance is ascribed to rectification of thermal noise in the drive nanowire during superconducting-normal transition.
    The suppression of the symmetric component is observed when a normal nanowire is used as either a drag or drive nanowire with the other nanowire superconducting. The absence of symmetric drag resistance between a normal drag nanowire and a superconducting drive nanowire suggests a higher electron-hole asymmetry in the superconducting LaAlO$_3$/SrTiO$_3$ nanowire arising from the 1D nature of superconductivity at LaAlO$_3$/SrTiO$_3$ interface.
 
	\end{abstract}
		
	\maketitle
    SrTiO$_3$ (STO)  has long attracted interest as a superconducting semiconductor \cite{schooley1964superconductivity,lin2013fermi,pai2018}. Recently, interest in the superconducting properties of STO was revived by the development of STO-based heterostructures and nanostructures and with the LaAlO$_3$/SrTiO$_3$ (LAO/STO) system \cite{Ohtomo2004} in particular. The LAO/STO two-dimensional interface supports superconductivity, which is electrostatically gateable, and various transport techniques have been used to study the superconductivity at the interface \cite{richter2013interface}.
  The superconducting transition temperature ($T_\text{C}$) has a dome shape as a function of carrier density, which is controllable via a backgate \cite{Caviglia2008}.
  With the use of conductive-atomic force microscope (c-AFM) lithography, nanoscale control over the conductance of the LAO/STO interface is possible.
  This technique relies on AFM tip-controlled protonation or deprotonation of the LAO surface, which enables the creation of a wide variety of quantum-confined structures, including superconducting nanowires \cite{Veazey2013}, ballistic 1D electron waveguides \cite{annadi2018quantized}, and single-electron transistors \cite{cheng2011sketched,Cheng2015}. These mesoscopic devices, drawn from a well-established toolset of quantum transport, often exhibit surprising new properties due to the unique physics of the STO interface such as electron pairing without forming superconductivity \cite{Cheng2015}. Recently by studying the superconductivity in LAO/STO nanowires of different widths and numbers, it is discovered that superconductivity exists at the boundary of nanowires and is absent within the interior region of nanowires, which indicates the 1D nature of superconductivity at LAO/STO interface \cite{pai2018}.
    
    Coulomb drag \cite{Narozhny2016}, or more generally frictional drag, first proposed by Pogrebinskii \cite{Progrebinskii1977}, has proven to be a powerful technique to study electron transport and electron correlations. When two electrical conductors are placed in close proximity, current driven through one (``drive") conductor may induce a voltage (or current) in the second (``drag") conductor. Frictional drag measurements have mostly been carried out between normal state conductors in coupled 2D semiconductor systems \cite{Gramila1991, Gramila1992, Gramila1994, Solomon1991, Eisenstein1992}, graphene systems \cite{Li2016,Lee2016}, 1D semiconductor systems \cite{Debray2001,Tokura2006,Laroche2014}, 1D complex oxide systems \cite{tang2019}, and quantum dot systems \cite{Keller2016}. 
Frictional drag in the superconducting regime has been carried out in normal metal-superconductor systems \cite{giordano1994cross, huang1995observation} and the phenomenon is explained by the local fluctuating electric field induced by mobile vortices in the superconducting layer \cite{shimshoni1995role} or Coulomb coupling between two conductors. \cite{kamenev1995coulomb, duan1993supercurrent}. 
    There are, to our knowledge, no prior reports of frictional drag between two quasi-1D superconductors.  
    
    Previously-reported frictional drag experiments at the LAO/STO-based nanowires have shown surprising results, particularly in the high magnetic field regime \cite{tang2019}. The drag resistance is anti-symmetric, indicating that the drag resistance arises via rectification of quantum shot noise in the drive nanowire due to the broken inversion symmetry of the drag nanowire \cite{Levchenko2008}. Remarkably, the drag resistance shows little to no dependence on the separation between nanowires (up to $\sim \mu$m scales). This unusual scaling strongly indicates that non-Coulombic interactions dominate the coupling between these nanowires. 
    
    Here we report frictional drag experiments between two LAO/STO superconducting nanowires. The drag resistance contains a mixture of symmetric and anti-symmetric components and the symmetric component disappears whenever one nanowire is normal and the other is superconducting.
    The antisymmetric component arises for the same reasons as in the high $B$ regime. The symmetric component is ascribed to the rectification of thermal noise in the drive nanowire during the superconducting-normal transition. Suppression of the symmetric drag component, when a normal nanowire is used as the drag nanowire, suggests the existence of a higher electron-hole asymmetry  \cite{narozhny2000mesoscopic} in the superconducting LAO/STO nanowires arises from the 1D nature of superconductivity at LAO/STO interface.
   
\begin{figure}
		\includegraphics[width=3.4in]{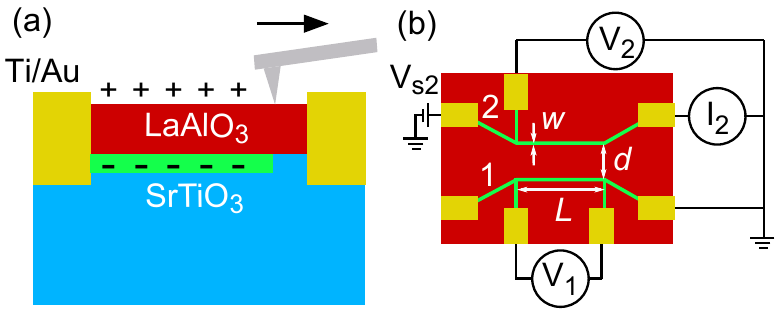}
		\caption{
			\label{Schematic}
Experimental setup. (a) Side-view of the nanowire fabrication process. A nanowire is created at the LAO/STO interface between two Ti/Au electrical contacts with c-AFM lithography. Protons ($+$) patterned on the surface by the AFM tip attract electrons ($-$) to the interface forming a nanowire (green area). (b) Top-view schematic of the double nanowire device with length $L$, width $w$, and nanowire separation $d$. The setup measures the induced drag voltage $V_1$ across nanowire 1 created by current $I_2$, which is induced by application of a voltage $V_{S2}$ across nanowire 2.
				}
\end{figure}

\begin{figure}
		\includegraphics[width=3.4in]{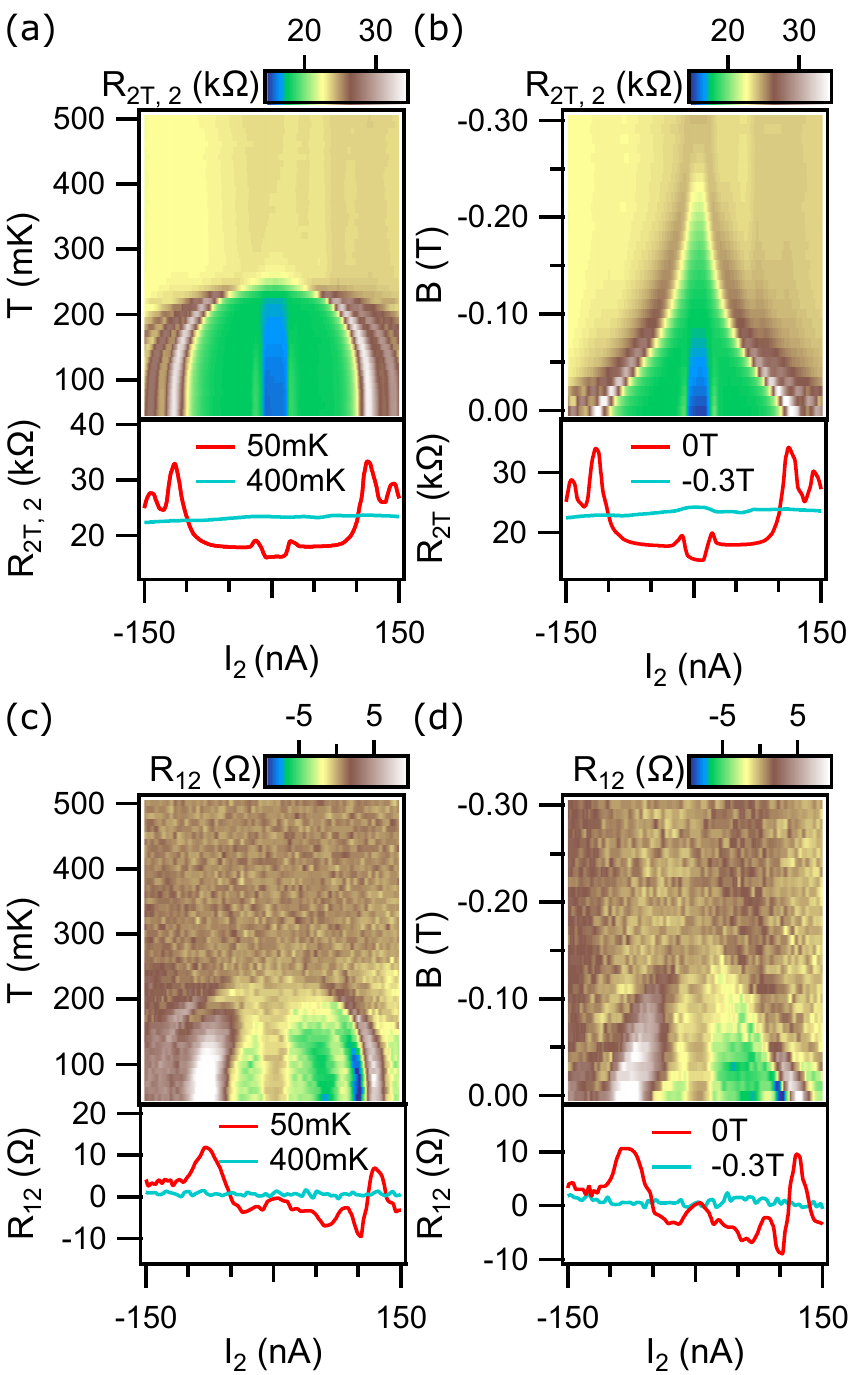}
		\caption{
			\label{Main data}
			Temperature and magnetic-field dependence of the drag resistance and two-terminal resistance of the drive nanowire. (a) Temperature dependence and line profiles of the drive nanowire's two-terminal resistance $R_{2T, 2}$ from nanowire 2. Top panel, temperature dependence of drag resistance $R_{12}$ from nanowire 1. Bottom panel, line profiles of $R_{12}$ at 50 mK and 400 mK. (b) Magnetic field dependence and line profiles of the drive nanowire's two-terminal resistance $R_{2T, 2}$. (c) Temperature dependence and line profiles of drag resistance $R_{12}$ from nanowire 1. (d) Magnetic field dependence and line profiles of drag resistance $R_{12}$.
		}
	\end{figure}

\begin{figure}
		\includegraphics[width=6.7in]{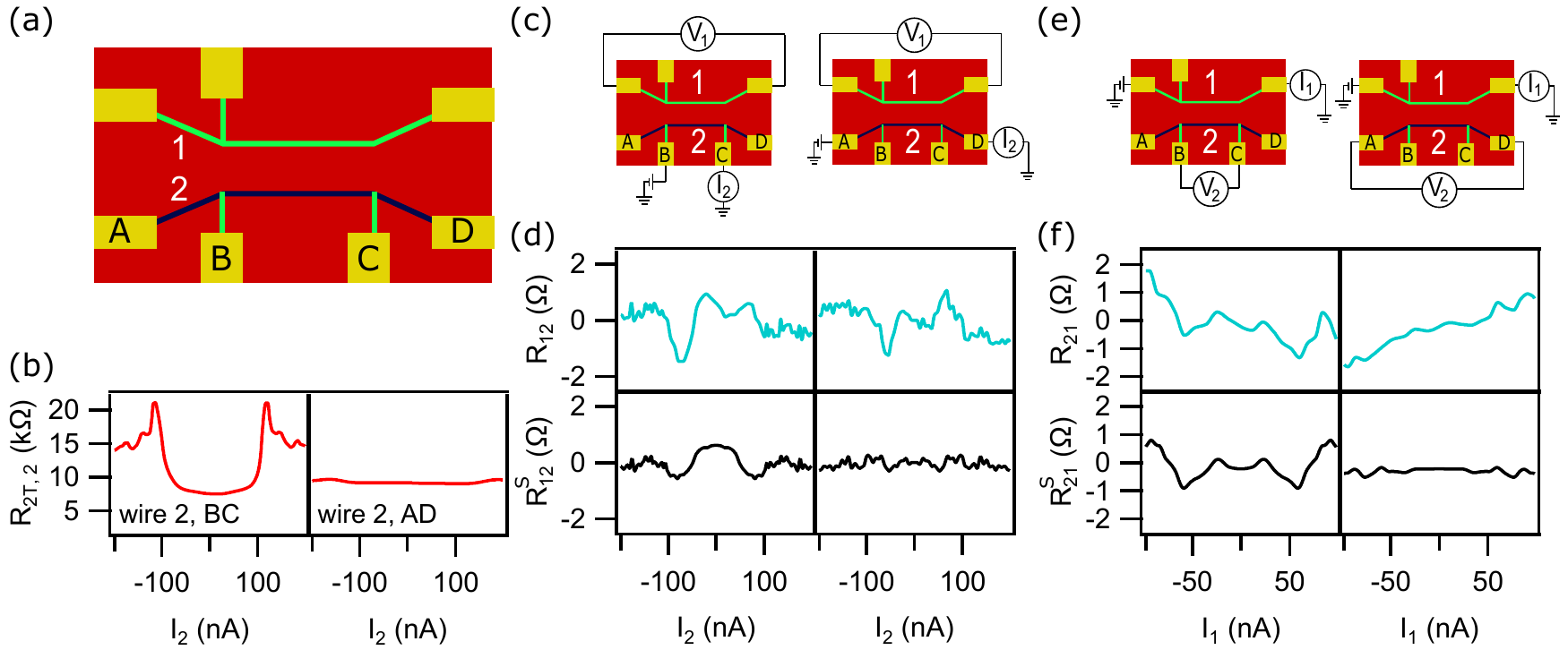}
		\caption{
			\label{Normal wire experiment}
			Frictional drag in the superconducting regime with one normal-state nanowire. (a) Schematic of the device with normal-state nanowire. Black sections in nanowire 2 are normal; green sections in nanowire 2 and 1 are superconducting. (b) Left: Two terminal resistance of nanowire 2 measured between B and C. Superconductivity arises from the green portions as shown in panel (a). Right: Two terminal resistance of nanowire 2 between A and D when the whole nanowire is in normal state. (c) Measurement configurations when nanowire 2 is used as the drive nanowire. (d) Drag resistance $R_{12}$ and its symmetric component $R^\text{S}_{12}$. Left and right panels correspond to the measurement configuration in (c). (e) Measurement configurations when nanowire 2 is used as the drag nanowire. (f) Drag resistance $R_{21}$ and its symmetric component $R^\text{S}_{12}$. Left and right panels correspond to the measurement configurations in (e).
		}
\end{figure}

   Nanowire devices are ``sketched" on LAO/STO heterostructures using c-AFM lithography \cite{ChengCen2008} (Fig. \ref{Schematic}(a)).  LAO/STO heterostructures with an LAO thickness of 3.4 unit cells are grown by pulsed laser deposition (PLD). Further details of the sample growth and the device fabrication process are described elsewhere \cite{WaterCycle}. The width of the nanowires used for these experiments is approximately $w=10$ nm, as quantified by erasure experiments \cite{ChengCen2008}. Other device parameters include the separation between nanowires $d$ and the nanowire length $L$. Here we focus on two sets of parameters: $d=40$ nm and $L=400$ nm (device 2B, Fig. \ref{Main data}) and $d=40$ nm and $L=300$ nm (device 2J, Fig. \ref{Normal wire experiment}). To investigate frictional drag at the LAO/STO interface in the superconducting regime, the magnitude of $B$ is kept below 0.3 T and and the temperature less than 100 mK (except for temperature-dependent measurements that explicitly go above $T=100$ mK). 
In a frictional drag experiment, a voltage $V_{i}$ in nanowire $i$ is induced by a current $I_{j}$ in nanowire $j$ (Fig. \ref{Schematic}(b)). All nanowires are connected to the same ground during the measurement. The current $I_{j}$ is produced by applying a voltage $V_{Sj}=V_\text{DC}+V_\text{AC}\cos\omega t$ to one end of nanowire $j$; the resulting current $I_{j}(\omega)$ and induced voltage $V_{j}(\omega)$ at frequency $\omega$ are measured using a lock-in amplifier. The resistance may then be expressed as a matrix $R_{ij}=dV_{i}/dI_{j}=V_{i}(\omega)/I_{j}(\omega)$, which is generally a function of the DC drive current $I_{j}$ (as well as other parameters such as temperature $T$ and applied magnetic field $\vec{B}$).
	The off-diagonal terms then define the drag resistance $R_{ij}$ characterize the mutual friction between electrons in the drive and drag nanowires.
	In order to ensure that the drag resistances $R_{ij}$ are not influenced by current leakage between the two nanowires, all measurements are performed well below the inter-wire breakdown voltage ($\sim$10 V) measured for each device.

    Typical frictional drag resistance measurements in the superconducting regime are shown in Fig. \ref{Main data}. Both nanowires in device 2B show signatures of superconductivity \cite{Veazey2013}. As shown in the bottom panels of Fig. \ref{Main data}(a) and (b), nanowire 2 displays three superconducting-normal transitions with critical current $I_\text{c}$ defined as the location of the peaks in $R_\text{2T,2}$ \cite{Veazey2013}. The first is at $\pm20$ nA, the second at $\pm110$ nA, and the third at $\pm140$ nA. Non-vanishing resistances in superconducting nanowires are common and are attributed to normal hotspots below $I_\text{c}$ \cite{tinkham2003hysteretic} or quantum phase slips \cite{giordano1994cross}.
    The superconducting-normal transition at $\pm20$ nA arises from the nanowire since it shows up both in $R_\text{2T, 2}$ and four-terminal resistance $R_\text{22}$ and the transition at $\pm110$ nA and $\pm140$ nA arises from wires connecting the nanowire and electrodes since it only shows up in $R_\text{2T,2}$ (Fig. \ref{Supplement 1}).
    The drag resistance $R_{12}$ is greatly enhanced in the superconducting regime, as can be seen by examining both the temperature-dependence (Fig. \ref{Main data}(c)) and the magnetic-field dependence (Fig. \ref{Main data}(d)).
   The nature of $R_\text{12}$ in the superconducting regime is qualitatively different from the high magnetic field regime (where the nanowires are not superconducting). In the high magnetic field regime, the drag resistance $R_{ij}$ is antisymmetric \cite{tang2019} with respect to the sourcing current, while the superconducting response is asymmetric with drive current. 
    The superconducting $R_{ij}$ is mostly symmetric between $I_\text{2}=\pm40$ nA with two tiny dips at $\pm10$ nA. As the magnitude of $I_\text{2}$ increases, an anti-symmetric component starts showing up in $R_{ij}$ and $R_{ij}$ becomes asymmetric.

	\begin{figure}
		\includegraphics[width=3.4in]{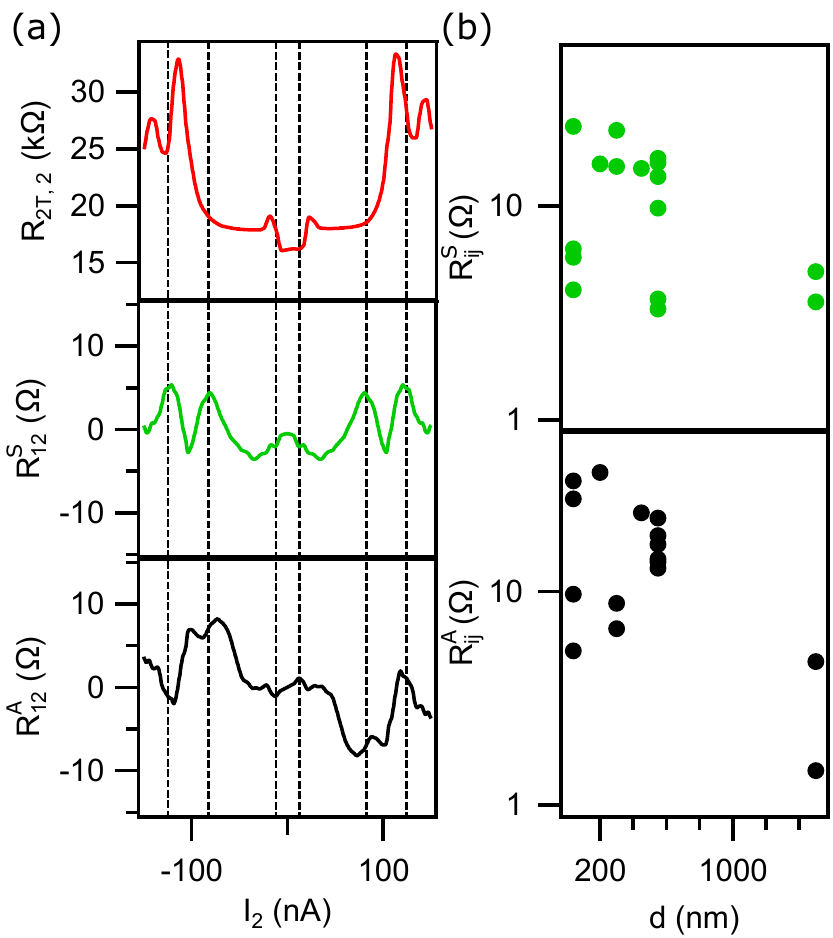}
		\caption{
			\label{Components}
			Symmetric and anti-symmetric components of drag resistance. (a) Typical symmetric and anti-symmetric components of drag resistance from device 2B. Top panel: Two-terminal resistance $R_\text{2T,2}$ of drive nanowire. Middle panel: Symmetric component of drag resistance $R_{12}$. Bottom panel: Anti-symmetric component of drag resistance $R_{12}$. (b) $d$ dependence of symmetric and anti-symmetric components. $d$ ranges from 40nm to 1.5$\mu$m. Top panel: Symmetric component $R^\text{S}_{ij}$ as a function of $d$. Bottom panel: Anti-symmetric component $R^\text{A}_{ij}$ as a function of $d$.
		}
	\end{figure}

The appearance of asymmetric $R_{12}$ (Fig. \ref{Main data}(c) and (d)) in the superconducting regime is correlated with the superconductivity in the drive nanowire 2 (Fig. \ref{Main data}(a) and (b)). To further understand the frictional drag in the superconducting regime, we extract symmetric and anti-symmetric components by $R_{ij}^\text{S}(I) = (R_{ij}(I) + R_{ij}(-I)) / 2$ and $R_{ij}^\text{A}(I) = (R_{ij}(I) - R_{ij}(-I)) / 2$. $R_\text{2T, 2}$, $R_{ij}^\text{S}$ and $R_{ij}^\text{A}$ are shown in top, middle an bottom panels of Fig. \ref{Components}(a). Dashed lines pinpoint locally strongest drag resistance in $R_{12}^\text{S}$.
As shown in Fig. \ref{Components}(a), locally strongest $R_{12}^{S}$ show up around the superconducting-normal transition represented by peaks in $R_\text{2T, 2}$ in the drive nanowire 2 accompanied by locally strongest $R_{12}^\text{A}$. The nature of coupling between nanowires for $R_{ij}^\text{S}$ and $R_{ij}^\text{A}$ is still unknown. But according to devices with $d$ ranging from 40nm to 1.5$\mu$m , both $R_{ij}^\text{S}$ and $R_{ij}^\text{A}$ persist over large separations and are nearly independent of $d$ (Fig. \ref{Components}(b)). Since the $e^{-4k_Fd}$ behavior is not observed in both $R_{ij}^\text{S}$ and $R_{ij}^\text{A}$, where $k_F\sim$(10nm)$^{-1}$ is the Fermi wave vector, the Coulomb coupling can be ruled out as the dominating effect \cite{Raichev2000}.

To corroborate that the symmetric component of drag resistance is related to the superconducting-normal transition in the drive nanowire, we examine the drag resistance from devices with one superconducting nanowire and one normal nanowire. The superconducting properties of LAO/STO is known to be gate-tunable both in 2D geometries \cite{Caviglia2007} and in 1D \cite{Veazey2013,pai2018}. There are known inhomogeneities in electron density which most likely arise from the underlying ferroelastic domain structure\cite{Aditi2019}.  While we cannot independently control the carrier density of one nanowire while keeping the second fixed, we can select devices in which one nanowire shows superconducting behavior and the other does not.
Fig. \ref{Normal wire experiment} shows the typical data from Device 2J. As illustrated in Fig. \ref{Normal wire experiment}(a), green-colored nanowires are superconducting, while black nanowires are in the normal-state. The information about the state of the nanowires is inferred from two-terminal resistance measurements (Fig. \ref{Normal wire experiment}(b)).
We then can compare the frictional drag as sensed by nanowire 1 due to two configurations--one in  which one device contains a superconducting section and one in which the other does not.

First we consider the configuration where superconducting nanowire 1 is the drag nanowire and examine the influence of drive nanowire's state on drag resistance, as shown in Fig. \ref{Normal wire experiment}(c). When both the drive and drag nanowires are superconducting, the drag resistance $R_{12}$ is asymmetric (Fig. \ref{Normal wire experiment}(d) left top panel) with a large symmetric component (Fig. \ref{Normal wire experiment}(d) left bottom panel). However, when the drive nanowire is normal, the drag resistance is mostly anti-symmetric (Fig. \ref{Normal wire experiment}(d) right top panel) with a negligible symmetric component (Fig. \ref{Normal wire experiment}(d) right bottom panel). 

The symmetric component of drag resistance showing up around the superconducting-normal transition in the drive nanowire can be explained by the rectification of the thermal noise in the drive nanowire \cite{Levchenko2008} which requires electron-hole asymmetry in both drive and drag nanowires.
When a superconducting nanowire undergoes superconducting-normal transition, the nanowire's resistance increases. This process generates thermal energy, which in turn gives rise to a large thermal noise and a greatly enhanced symmetric component of drag resistance. For the normal nanowire, therefore there is no significant enhancement of the thermal noise, and the symmetric component of drag resistance remains small at all biases across the drive nanowire.

The rectification of thermal noise in the drive also explains the strong correlation between $R_{12}^\text{A}$ and $R_{12}^\text{S}$. $R_{12}^\text{A}$ comes from the rectification of the shot noise in the drive nanowire \cite{tang2019}. Shot noise is a non-equilibrium phenomenon depending on the voltage bias across the drive nanowire \cite{Levchenko2008}. During the superconducting-normal transition in the drive nanowire, the change of drive nanowire's resistance changes the bias across different portions of the nanowire, thus inducing quantum shot noise and anti-symmetric $R_{ij}$ is observed simultaneously with symmetric drag resistance. 

The symmetric component in drag resistance is also strongly suppressed when the drag nanowire is in the normal state. As shown in the left panel of Fig. \ref{Normal wire experiment}(e), when the drag resistance is measured between B and C of nanowire 2, the drag resistance $R_{21}$ is asymmetric with a large symmetric component (Fig. \ref{Normal wire experiment}(f) left). However when the drag resistance is measured between A and D, the drag resistance is anti-symmetric with a negligible symmetric component.
Since the drive nanowire 1 is superconducting in both configurations, the absence of symmetric drag resistance component with a normal drag nanowire cannot be ascribed to the absence of thermal noise in the drive nanowire. The magnitude of symmetric drag resistance depends on the electron-hole asymmetry in the drag nanowire \cite{Narozhny2016}. The fact that the symmetric drag resistance measured from a superconducting drag nanowire is larger may be explained by the electron-hole asymmetry is stronger in superconducting nanowire than normal nanowire. Electron-hole symmetry is more easily broken in low dimensional devices \cite{narozhny2000mesoscopic, Levchenko2008}. It is reported that the superconductivity at LAO/STO interface is 1D in nature situated at the boundary of the nanowire and is absent within the interior region of the nanowire \cite{pai2018}. Thus the overall dimension of the nanowire is reduced as it becomes superconducting compared to a normal nanowire due to the formation of 1D superconducting boundary. This reduced dimension of the nanowire gives rise to a stronger electron-hole asymmetry. Therefore the symmetric component of drag resistance is stronger measured from a superconducting drag nanowire.

In summary, frictional drag between superconducting LAO/STO nanowires exhibits a strong and highly symmetric component in drag resistance, which is distinct from the anti-symmetric drag resistance between LAO/STO nanowires in the normal state. The symmetric component arises from rectification of thermal noise in the drive superconducting nanowire based on the fact that it shows up at the vicinity of superconducting-normal transition in the drive nanowire and disappears when the drive nanowire is normal. The symmetric component in drag resistance also disappears when the drag nanowire is normal, which can be attributed to the 1D nature of superconductivity in LAO/STO systems.

\begin{acknowledgements}
Work at the University of Pittsburgh was supported by funding from the DOE Office of Basic Energy Sciences under award number DOE DE-SC0014417. The work at the UW-Madison (synthesis and characterizations of thin films heterostructures) was supported by the US Department of Energy (DOE), Office of Science, Office of Basic Energy Sciences (BES), under award number DE-FG02-06ER46327.
\end{acknowledgements}

\bibliography{SCFrictionaldrag}

\pagebreak

\begin{center}
\textbf{\large Supplemental Material}
\end{center}
\onecolumngrid
\renewcommand{\thefigure}{S\arabic{figure}}
\setcounter{figure}{0}

    In the supplement materials, Fig. \ref{Supplement 1} shows the typical two-terminal ($R_\text{2T, 2}$) and four-terminal resistance ($R_\text{22}$) from a nanowire. The superconducting-normal transition at small bias showing up in both two-terminal and four-terminal resistance comes from the nanowire. The extra superconducting-normal transitions at larger biases in $R_\text{2T, 2}$ come from wires connecting the nanowire and the electrode.
    
	\begin{figure}[h]
		\includegraphics{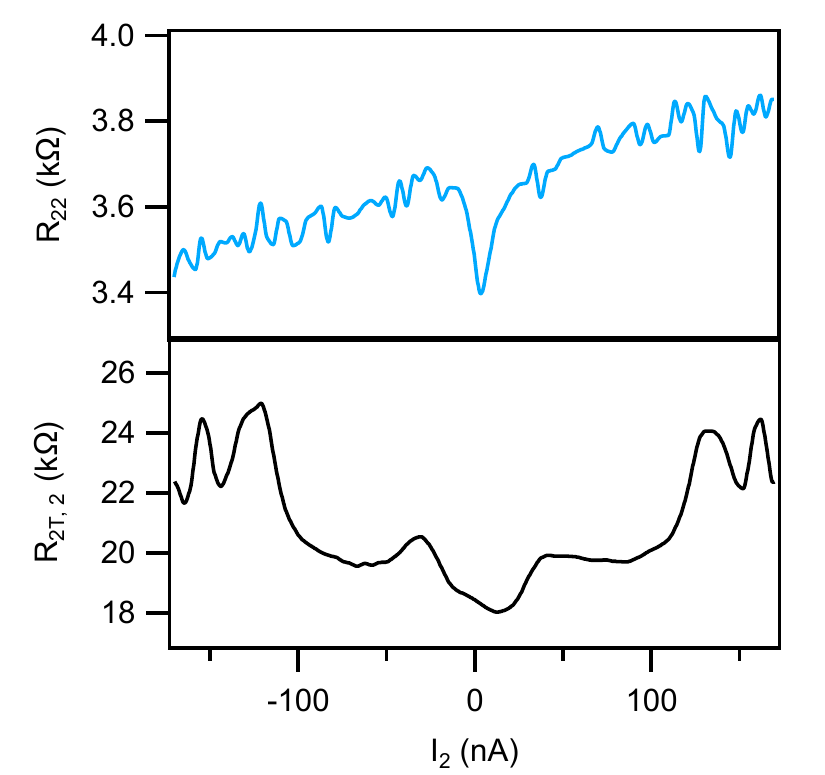}
		\caption{
		\label{Supplement 1}
			Typical two-terminal resistance and four-terminal resistance from a nanowire. Top panel: Four-terminal resistance $R_\text{22}$ and superconducting-normal transition from the nanowire only shows up at small bias from $\pm20$ nA. Bottom panel: Two-terminal resistance $R_\text{2T, 2}$. Besides the superconducting-normal transition at small bias extra superconducting-normal transitions show up at larger bias $\pm110$ nA and $\pm150$ nA. Superconducting-normal transitions at larger bias come from wires connecting the nanowire and electrodes.
		}
	\end{figure}
	 
\end{document}